\documentclass[prb,preprint,superscriptaddress,nobibnotes,endfloats*]{revtex4}

\usepackage{graphicx}
\usepackage{dcolumn}
\usepackage{bm}
\usepackage{times,mathptmx}
\usepackage{color}
\usepackage{multirow}
\usepackage{enumitem}
\usepackage{chngpage}
\usepackage{amsmath}

\begin{document}


\title{\bf Chip-scale cavity optomechanics in lithium niobate}

\author{Wei C. Jiang}
\affiliation{Institute of Optics, University of Rochester, Rochester, NY 14627}
\author{Qiang Lin}
\email{qiang.lin@rochester.edu}
\affiliation{Institute of Optics, University of Rochester, Rochester, NY 14627}
\affiliation{Department of Electrical and Computer Engineering, University of Rochester, Rochester, NY 14627}

\date{\today}


\begin{abstract}
We develop a chip-scale cavity optomechanical system in single-crystal lithium niobate that exhibits high optical quality factors and a large frequency-quality product as high as $3.6\times 10^{12}$ Hz at room temperature and atmosphere. The excellent optical and mechanical properties together with the strong optomechanical coupling allow us to efficiently excite the coherent regenerative optomechanical oscillation operating at 375.8 MHz with a threshold power of 174~${\rm \mu W}$ in the air. The demonstrated lithium niobate optomechanical device enables great potential for achieving electro-optic-mechanical hybrid systems for broad applications in sensing, metrology, and quantum physics.

\end{abstract}

\maketitle

Cavity optomechanics \cite{Marquardt14} has emerged as a fascinating field exploring the coherent interaction between light and mechanical motion mediated by radiation pressure. Facilitated by photonic micro-/nano-cavities with greatly enhanced optomechanical coupling, diverse chip-scale optomechanical systems have been developed on a variety of material platforms \cite{Vahala05, Painter09, Favero10, Wong10, Kippenberg11, Jiang12, Tang12, Pernice13, Barclay15, Kartik15, Lin15}, showing great promise for broad applications ranging from precision measurements and sensing \cite{Kippenberg09, Kippenberg12, Painter12}, to frequency metrology \cite{Vahala06, Bhave11,Luan14}, information processing \cite{Tang11, Painter12_2, Wang12}, and quantum physics \cite{Painter11, Teufel11, Painter11_2, Painter11_3, Bochmann13, Kartik16}. In particular, materials such as aluminium nitride \cite{Tang12, Bochmann13}, diamond \cite{Pernice13, Barclay15, Barclay15_2, Loncar15}, and silicon carbide \cite{Lin15}, have recently attracted great research interest for cavity optomechanics, due to their wide electronic band gap to allow broadband optical operation, large refractive index to enable strong optical confinement, and excellent mechanical properties to achieve high mechanical frequency and low mechanical dissipation.

Lithium Niobate (LN) \cite{Gaylord85}, well known for its unique features exhibiting large electro-optic, nonlinear-optical, acousto-optic, and piezoelectric effects, has been widely used in optoelectronics, nonlinear photonics, quantum photonics, and microelectromechanical systems (MEMS) with a variety of applications ranging from high-speed optical modulators, optical parametric oscillators, and entangled photon sources, to acoustic wave transducers, sensors, and filters.  Recent advances in single-crystal LN thin film technologies \cite{Levy98, Gunter12} enable strong light confinement in sub-micron structures to realize micro-/nano-photonic high-Q cavities and resonators \cite{Gunter07, Niu12, Diziain13, Loncar14, Cheng15, Xu15}. A combination of its large electronic bandgap (4 eV) and outstanding optical and mechanical properties makes LN an excellent material platform for cavity optomechanics, and it could potentially offer a bridge for combining photonic, electronic, and mechanical components into an integrated hybrid system with versatile functionalities. 

In the present work, we demonstrate the first LN cavity optomechanical system realized on lithium-niobate-on-insulator (LNOI), which exhibits high optical Q in telecom wavelengths, low mechanical dissipation in atmosphere, and high mechanical frequency above the VHF band. By taking advantage of both high optical and mechanical qualities of our device, we successfully demonstrate regenerative optomechanical oscillation in the air.

\begin{figure}[h]

 \centering\includegraphics[scale=0.38]{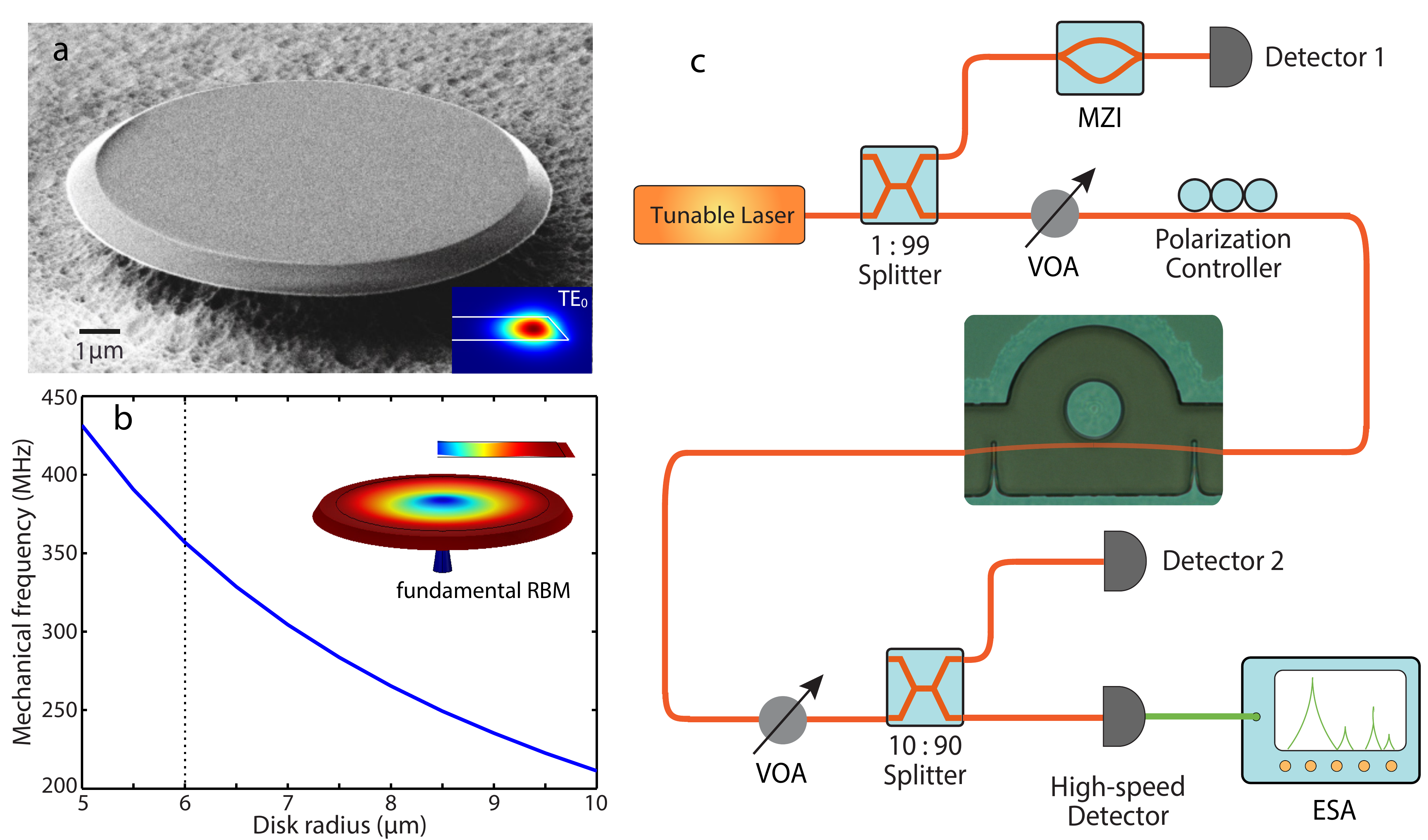}
\caption{(a) SEM image of a fabricated LN microdisk with a radius of 6 ${ \rm \mu m}$. The slant angle of the sidewall is around 45 degree. Inset: simulated optical mode profile of the fundamental quasi-transverse-electric (quasi-TE) mode by FEM. (b) Simulated mechanical frequency of the fundamental RBM as a function of radius for the LN microdisk with a thickness of 400 nm. Inset: Simulated mechanical displacement profile of the fundamental RBM by FEM (For LN, Young's modulus ${E}$ = 181 GPa and mass density ${\rho}$ = 4640 kg/m${^3}$). (c) Experimental setup for optomechanical measurement. MZI:  Mach-Zehnder interferometer; VOA: variable optical attenuator; ESA: electrical spectrum analyzer. }
\label{MechMode}
\end{figure}

\begin{figure}[h]

 \centering\includegraphics[scale=0.45]{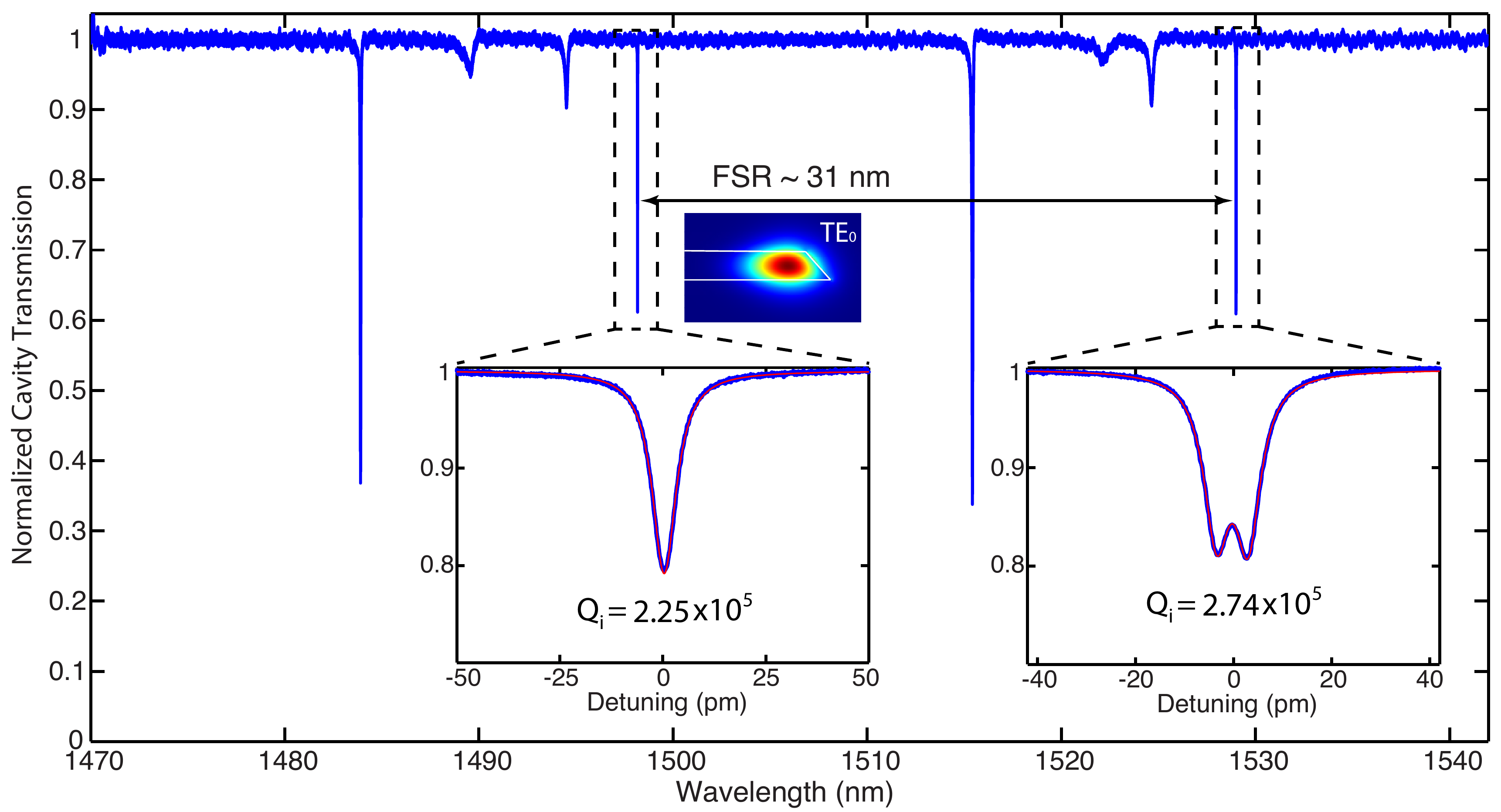}
\caption{The cavity transmission of the LN microdisk. The insets show detailed transmission spectra for the fundamental quasi-TE modes at 1498 nm and 1529 nm, respectively, with a theoretical fitting (red).}
\label{OpticalQ}
\end{figure}

The employed device structure for LN optomechanical system is microdisk resonators which support high-Q whispering gallery modes (inset of Fig.~\ref{MechMode}a).  We have developed a robust nano-fabrication technique (see Appendix) to realize high-Q LN microdisks. Fig.~\ref{MechMode}a shows a typical example of our fabricated microdisks with a radius of 6~${\rm \mu m}$ and device thickness of 400 nm sitting on a 2 ${\rm \mu m}$ thick oxide pedestal. The fabrication process is optimized to produce a smooth device sidewall, which is critical for minimizing the scattering loss of the optical modes to achieve high optical Q.

Radiation pressure induced by the optical mode in the device couples strongly to the mechanical radial-breathing modes (RBMs) (inset of Fig.~\ref{MechMode}b) whose dominant motion is along the radial direction of the microdisk. Simulation by the finite-element method (FEM) indicates the frequency of the fundamental RBM scales inversely with the device radius. Fig.~\ref{MechMode}b shows that shrinking the device radius below 7~${\rm \mu m}$ is able to realize a high mechanical frequency above the VHF band ($>$ 300 MHz). In particular, the 6-${\rm \mu m}$ radius microdisk exhibits a mechanical frequency of $\frac{\rm \Omega_m}{2\pi}=$357 MHz with an effective motional mass of 152 pg. Another advantage of a smaller microdisk is the larger optomechanical coupling coefficient, which scales as $\rm g_{om}= - \omega_0/R $, where $\omega_0$ is the optical resonance frequency. The detailed FEM simulation shows that the microdisk with a radius of 6 ${\rm \mu m}$ exhibits ${\frac{\rm |g_{om}|}{2\pi} = \sim}$ 28 GHz/nm for a telecom-band wavelength, which thus corresponds to a vacuum optomechanical coupling rate of $\rm \frac{ |g_0|}{2\pi} = \frac {|g_{om}|}{2\pi} \sqrt{\hbar/(2m_{eff}\Omega_m)}=\sim 11.2~kHz$ for the interaction between the fundamental quasi-transverse-electric (quasi-TE) mode and the fundamental RBM.

The fabricated 6-${\rm \mu m}$ radius microdisk is characterized with the experimental setup (shown in Fig. \ref{MechMode}c) at room temperature and atmosphere (see Appendix). Figure \ref{OpticalQ} shows the optical transmission spectrum of the LN microdisk from 1470 to 1545 nm, which clearly exhibits different mode families of quasi-TE modes with the lowest three radial orders. The fundamental quasi-TE modes with a free spectral range (FSR) of ${\sim}$ 31 nm exhibit high optical Q over the telecom band. The detailed cavity transmission spectra with Lorentzian fitting (insets of Fig.~\ref{OpticalQ}) show high intrinsic optical Qs of  $2.25\times 10^5$ and  $2.74\times 10^5$ for the fundamental quasi-TE modes at 1498 nm and 1529 nm, repectively.

\begin{figure}[htbp]
 \centering\includegraphics[scale=0.45]{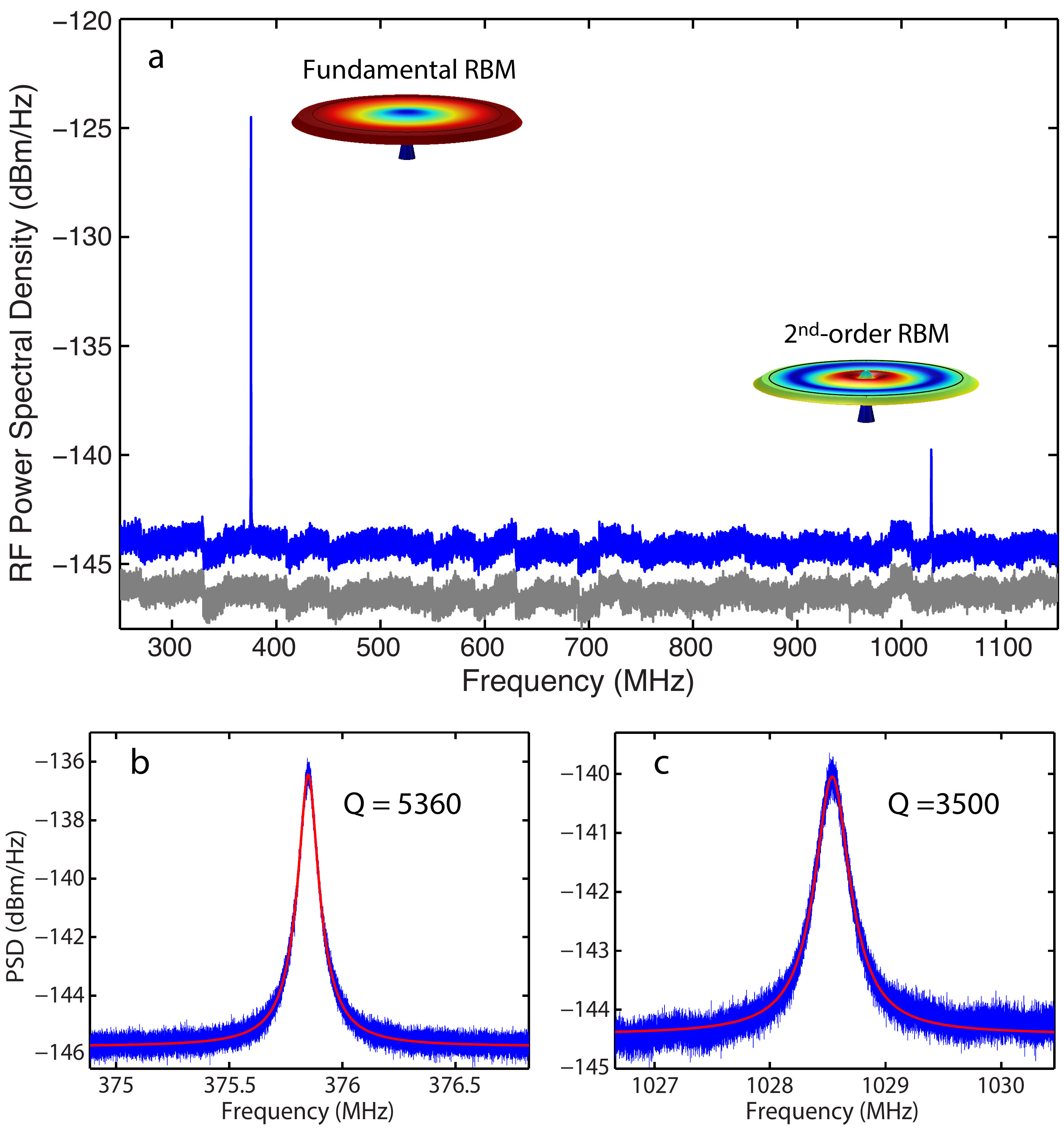}
\caption{(a) Optically transduced RF spectrum (blue curve) of the cavity transmission showing the fundamental and second-order RBMs at 375.8 MHz and 1.029 GHz, respectively. The gray trace shows the noise background of the high-speed photo-detector. (b) and (c) Detailed thermal noise spectra with Lorentzian fitting for the fundamental and second-order RBMs, respectively.}
\label{TBM}
\end{figure}

To characterize the mechanical quality of the device, we tune the input laser to the optical cavity mode at 1498 nm and record the radio-frequency (RF) spectrum of the optical transmission. To enhance the optical transduction of mechanical motion, we increase the coupling depth of the cavity transmission to realize a loaded optical Q of   $1.88\times 10^5$ and lock the laser frequency half way into the cavity resonance at the blue-detuned side. Fig. \ref{TBM}a shows the transduced RF spectrum of the mechanical modes for the microdisk with negligible optomechanical dynamic backaction effect at low optical drop power of a few microwatts. We can clearly observe the fundamental and second-order RBMs at 375.8 MHz and 1.029 GHz, respectively, which are within 10{\%} of the simulated values. Fig.~\ref{TBM}b and c show their detailed thermal noise spectra with Lorentzian fitting indicating that they have intrinsic mechanical linewidths of 70 and 294 kHz, respectively, resulting in intrinsic mechanical Qs of 5360 and 3500 while the device resides in the air. The high mechanical Qs result from achieving a minimal radius ($<$ 100 nm) of the supporting silica pedestal to dramatically suppress the clamping loss. Consequently, the mechanical modes exhibit frequency-quality ($f\cdot Q$) products of 
  $2.0\times 10^{12}$ and  $3.6\times 10^{12}$ Hz, which are the highest among the recent reported LN mechanical microresonators \cite{Piazza13, Olsson14, Bhave15}.

\begin{figure}[htbp]
 \centering\includegraphics[scale=0.52]{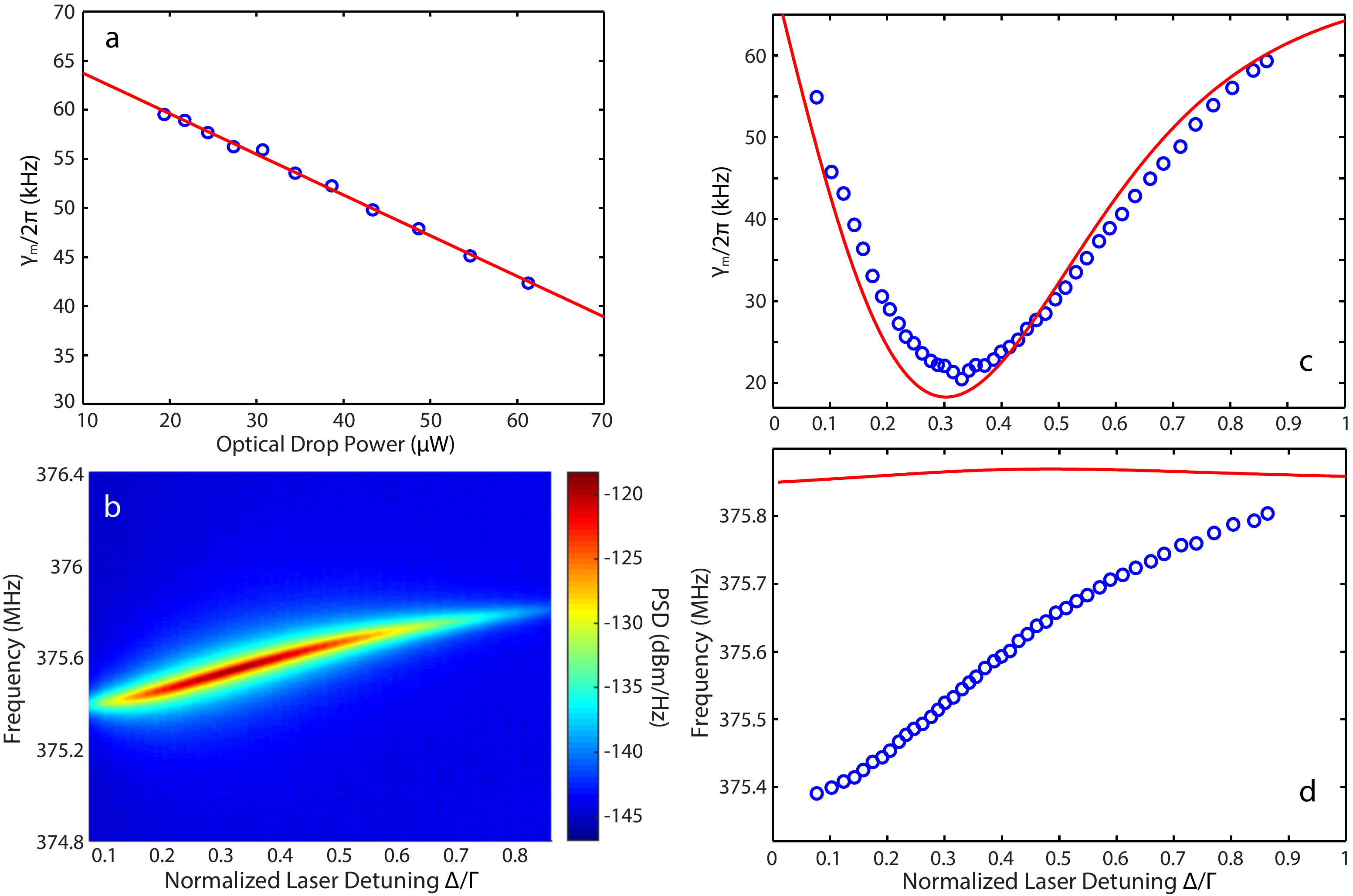}
\caption{(a) The mechanical damp rate as a function of optical drop power with a linear fitting in red. (b) Intensity image of the measured RF spectrum versus normailzed laser detuning at the blue-detuning side. Measured (c) mechanical damp rate and (d) mechanical frequency versus normalized laser detuning compared with the theoretical curves in red.}
\label{Backaction}
\end{figure}

When the optical power increases at the current detuning condition, the optomechanical backaction becomes pronounced resulting in decreasing the mechanical damp rate of the fundamental RBM as shown in Fig.~\ref{Backaction}a. By linear fitting, we can extract the optomechanical coupling rate $\rm \frac{ |g_0|}{2\pi} = 11.6~kHz$, which agrees with the simulated value. To further investigate the optomechanical backaction, we record the RF spectrum (Fig.~\ref{Backaction}b) by sweeping the laser frequency at the blue-detuned side while the input power is fixed at 330 ${\rm \mu W}$, which clearly shows the amplification of the mechanical motion and optical tuning of the mechanical frequency. Fig.~\ref{Backaction}c shows the measured mechanical damp rate as a function of the normalized laser detuning, which agrees with the theoretical curve quite well. 
The observed optimal detuning for optomechanical amplification is at $\rm \Delta \approx 0.33\Gamma$ resulting in a much narrowed mechanical linewidth close to 20 kHz, where $\Gamma$ is the photon decay rate of the loaded cavity. On the other hand, Fig.~\ref{Backaction}d shows that the measured mechanical frequency consistently decreases while the detuning approaches zero. Compared with the theoretical curve showing a slight frequency change due to the pure optical spring effect, the significant discrepency is due to the static thermal softening resulting from a reduction in the Young's modulus of LN as the device temperature increases \cite{Smith71}.

\begin{figure}[htbp]
 \centering\includegraphics[scale=0.4]{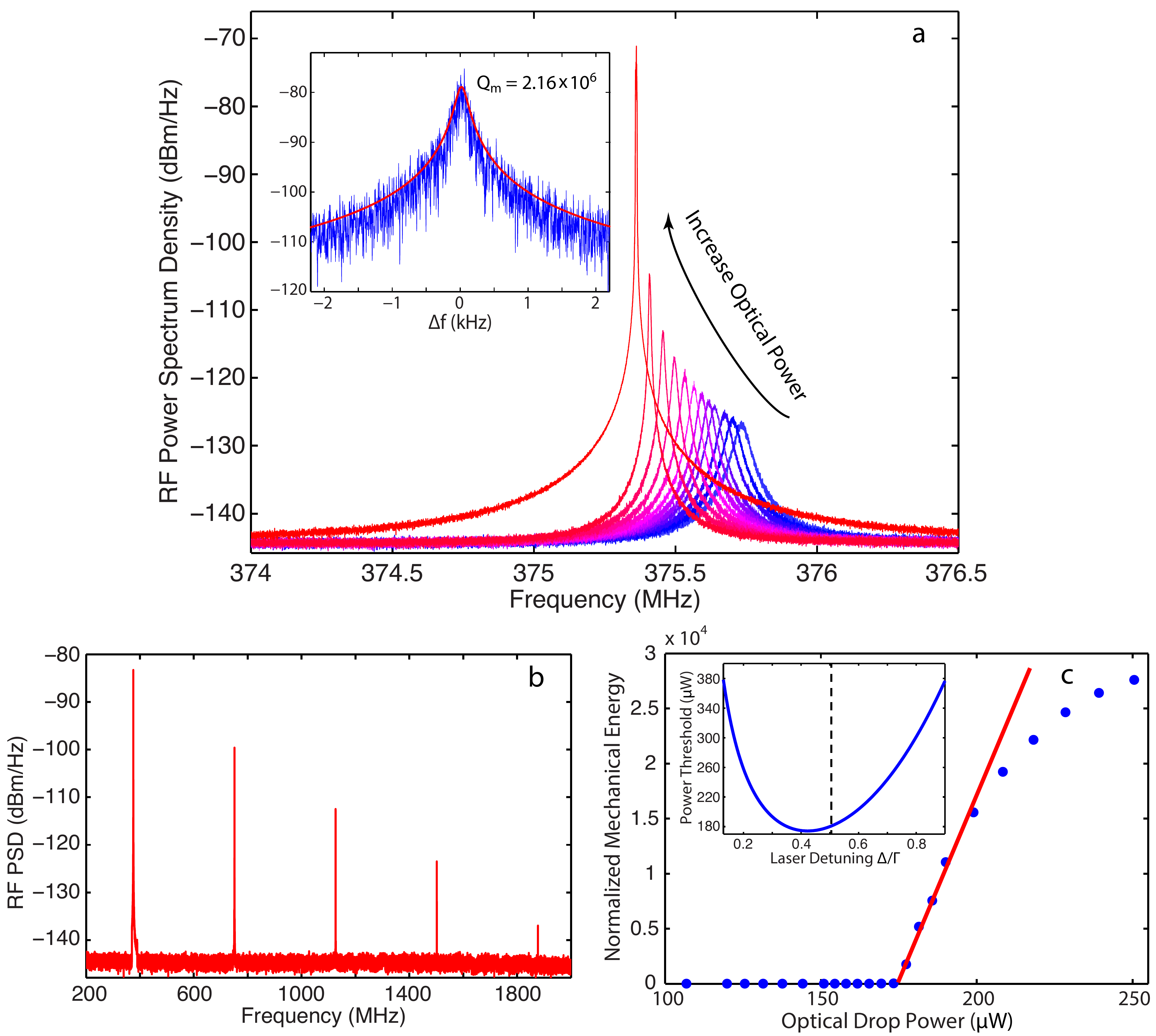}
\caption{(a) RF Spectrum of the fundamental RBM at various optical drop powers showing optomechanical oscillation. The inset shows the detailed oscillation spectrum with a Lorentzian fitting (red). (b) The self-oscillation spectrum exhibiting generation of a series of harmonics. (c) The normalized mechanical energy as a function of optical drop power. The red line is a linear fit to data above threshold. The inset shows the theoretical drop power threshold for our device as a function of normalized laser detuning. The dashed line indicates the detuning for the measurement.}
\label{PhononLaser}
\end{figure}

The strong dynamic backaction in the LN microdisk implies a correspondingly efficient amplification of the mechanical motion on the blue-detuned side of the cavity resonance, resulting in regenerative optomechancial oscillation. As shown in Fig.~\ref{PhononLaser}a, the RF spectrum of the fundamental RBM increases dramatically with increased input power, accompanied by a significant narrowing of the mechanical linewidth and a slight red shift in frequency. When the optical drop power is increased to 200 ${\rm \mu W}$, the peak value of the mechanical spectral intensity is dramatically enhanced by more than 50 dB. Detailed analysis of the oscillation spectrum (inset of Fig.~\ref{PhononLaser}a) shows that the mechanical linewidth is drastically suppressed down to 173 Hz, corresponding to an effective mechanical Q factor as high as $2.16\times 10^6$.  Fig.~\ref{PhononLaser}b shows the oscillation spectrum with a much broader span, clearly exhibiting generation of a series of harmonics up to the fifith order, which is limited by the bandwidth of the high speed photo-detector.

By integrating the spectral area of the transduced mechanical spectrum at various drop powers, we obtain the curve shown in Fig.~\ref{PhononLaser}c, where the mechanical energy is normalized by the room-temperature thermal mechanical energy in the absence of dynamic backaction. It shows a clear phonon-lasing behavior where the mechanical energy remains rather small when the drop power is below a certain level but starts to increase dramatically and depends linearly on the drop power above the threshold. By linear fitting the above-threshold data, we obtain a threshold power of 174~${\rm \mu W}$, which agrees with the theoretical value (see the inset of Fig.~\ref{PhononLaser}c). As the drop power increases further higher, the mechanical energy becomes saturated since the large mechanical displacement induces the cavity frequency shift larger than the cavity linewidth. 

In summary, we have demonstrated chip-scale cavity optomechanics in lithium niobate for the first time, and the high performance of the demonstrated device shows great potential for sensing and metrology applications. The current optical Q is still limited by the optical scattering loss and can be possibly further improved to above $10^6$ either by optimizing the fabrication process or by using a laser with a shorter wavelength. Furthermore, by scaling down the dimension of the microdisk (R $< \rm 2~\mu m$) or utilizing the optomechanical crystal structures \cite{Painter09}, giant vacuum coupling rates and resolved-sideband regime can be achieved which are favorable for demonstration of various phenomena in quantum physics, such as quantum ground state cooling, optomechanically induced transparency, and squeezing of light and nanomechanical motion. 

Moreover, the demonstrated LN optomechanics provides a new route to realizing integrated electro-optic-mechanical hybrid systems in the near future. By taking advantage of the excellent piezoelectric properties of LN, we would be able to achieve a monolithic nanomechanical interface between microwave signals and optical photons enabling translation of quantum states \cite{Bochmann13, Kartik16}. Finally, incorporation of high-fidelity photon sources and electro-optic tunability on LN would create a new paradigm for integrated quantum photonics with diversified functionalities and provide great potential for broad applications in quantum communication, quantum computing and network.

\section*{Appendix}

\subsection{Device fabrication}\label{Device_fab}

The microdisk resonator is fabricated from a single-crystal lithium niobate (Z-cut) on insulator wafer (by NANOLN), with a top device layer of 400 nm thickness and a buried oxide layer of 2 ${\rm \mu m}$ thickness on a LN substrate. The detailed process procedure is described as follows. First,  ZEP-520A positive electron-beam resist is spin-coated onto the wafer with a thickness of $\rm \sim 1 \mu m$, and is then patterned by electron-beam lithography (EBL) system JEOL9500. The developed pattern is transferred to the LN layer via Ar-ion mill etching resulting in smooth and slant sidewalls. Finally, the buried oxide layer is isotropically etched by diluted hydrofluoric (HF) acid with precise control to form a supporting pedestal (radius $\rm < 100~nm$).

\subsection{Experimental setup}\label{EXP}

The experimental setup (Fig. \ref{MechMode}c) is described as follows. A continous-wave tunable laser with a wavelength range of 1470-1545 nm is launched into the device by evanescent coupling via a fiber taper which is docked onto two nanoforks for stable operation. Before the device, a variable optical attenuator (VOA) is used to adjust the input optical power and a polarization controller is used to select the polarization of the incident wave. The transmitted light is first detected by a slow photo-detector to record the cavity transmission to characterize the optical Q, which is calibrated by a Mach-Zehnder interferometer (MZI). Then the wave transmitted from the cavity carrying the information of mechanical motion is detected by a high-speed photo-detector whose output is characterized by an electrical spectrum analyzer (ESA). The device is tested in the air environment at room temperature.

\section*{Acknowledgements}

This work was supported by National Science Foundation under Grant No. ECCS-1509749. It was performed in part at the Cornell NanoScale Science $\&$ Technology Facility (CNF), a member of the National Nanotechnology Infrastructure Network.



\end{document}